\documentclass[copyright,creativecommons]{eptcs}
\providecommand{\event}{WCSI 2010} 
\usepackage{breakurl}             

\usepackage{color}
\usepackage{epsfig}
\usepackage{wrapfig}
\usepackage{pstricks} 
\usepackage{fancyvrb}
\usepackage{pslatex} 
\usepackage{listings} 
\usepackage{subfig}  
\usepackage{floatflt}
\usepackage{amsmath,amssymb,amsfonts}
\usepackage{multirow}
\usepackage{epsfig}
\newcommand{\pa}{\emph{ProActive}}

\newtheorem{Definition}{Definition}

\def\ignore#1{}

\newcommand{\symb}[1]{\makebox{\it #1}}

\newcommand{\fv}{\symb{fv}}

\DeclareMathAlphabet{\drawsfont}{OT1}{cmss}{m}{n}
\DeclareMathAlphabet{\drawsbold}{OT1}{cmss}{bx}{n}

\newcommand{\bs}{$\hspace{-.2ex}$}

\title{Behavioural Models for Group Communications}

\author{Rab\'ea Ameur-Boulifa 
\institute{System-on-Chip Laboratory (LabSoC),
Telecom Paristech, Sophia Antipolis, France}
\email{Rabea.Ameur-Boulifa@telecom-paristech.fr}
\and
 Ludovic Henrio and Eric Madelaine
\institute{INRIA, CNRS, I3S, University of Nice Sophia-Antipolis,
  Sophia Antipolis, France}
\email{Ludovic.Henrio@sophia.inria.fr and Eric.Madelaine@sophia.inria.fr}}
\def\authorrunning{R. Ameur-Boulifa, L. Henrio, and E. Madelaine}
\def\titlerunning{Behavioural Models for Group Communications}

\begin{document}

\maketitle

\begin{abstract} 
Group communication is becoming a more and more popular infrastructure for
efficient distributed applications. It consists in representing
locally a group of remote objects as a single object accessed in a single
step; communications are then broadcasted to all members.
This paper provides  models  for automatic verification of
group-based applications,  typically for detecting deadlocks or
checking message ordering. We show how to encode
group communication, together with different forms of synchronisation
for group results.
The proposed models are parametric such that, for example, different
group sizes or group members could be experimented with the minimum
modification of the original model. 
\end{abstract}

\section{Introduction}

Group communication is a communication pattern allowing a single
process to perform a communication to many clients in a single
instruction, this operation can be synchronized or optimized accordingly.
Nowadays group communication  is widely used in distributed computing particularly in grid technologies \cite{Grid05}.
Objects can register
to a group and receive communications handled in a collective
way. 
Group membership is transparent to the receiver that simply handles
requests it receives. Group communications are also easy to handle on
the sender side because a simple invocation can trigger several
communications. Communication parameters are sent according to a
distribution policy; they can be for example broadcasted or split
between the members of the group. Several  middleware platforms and toolkits for building distributed applications implement  one-to-many communication mechanisms
\cite{JGroup,baduel07asynchronous,KPS+04}.

This paper addresses the crucial point of reliability of distributed
applications using group communications. The most frequent reliability
issue for distributed application is to be able to detect deadlocks,
in the case of group, a dead lock can occur for example when a member
of the group does not answer to its requests while the request sender is waiting for all the results. Such an absence of response might be due to an issue in message ordering for example.
In order to enhance reliability of group applications we  develop
methods for the analysis and  verification of behavioural properties
of  such  applications, our method can be applied with automatic tools.

A first contribution of this paper is to provide a model allowing the
verification of the behaviour of group-based applications, in other
words, we provide a verifiable model for group communication. We also
illustrate our approach by specifying an application example,
instantiating the verifiable model, and proving a few properties.

To precisely define the semantics of group communications, we focus
on  a specific
middleware called \pa~\cite{Pro03}. \pa\ provides  a high-level programming API for building
distributed applications, ranging from Grid computing to mobile
applications. 
\pa\ offers advanced communication strategies, including group
communication \cite{LSR-ARTICLE-2006-001,baduel07asynchronous}.
In \pa, remote communication relies on asynchronous requests
with futures: upon a call on a remote entity, a request is created at
the receiver side, and a future is created on the sender side
that will be filled when the remote entity provides an answer.
What make the handling of groups particular in \pa\ is the necessity
to also gather and manage replies for requests sent to the group.
Synchronisation on futures is generally transparent: an access
to a future blocks until the result is computed and returned. However,
synchronisation on group of futures, that represent the result of a
group invocation,
features more specific and complex synchronization primitives. Consequently, our model also
encodes different synchronisation policies.


In \cite{BBCHM:article2009} we have defined a parameterized and hierarchical model for synchronised networks of labelled transition systems. We have shown how this model can be used as an intermediate format to represent the behaviour of distributed applications, and to check their temporal properties.
In this paper, we present a method for building parameterized models
capturing the behavioural semantics of group communication systems;
models are the networks of labelled transition systems, whose labels
represent method invocations. The language we chose is pNets; it is an
intermediate language: the models we present here should be generated,
either from source code or from a higher-level specification. PNets
themselves are then used to generate a model in a lower-level language
that will be used for verification of the program properties.
In this paper the advantage of choosing such an intermediate language
are the following: compared to a higher-level
language, pNets are precise enough to define a behavioural semantics,
and compared to lower level languages, they provide parameterized
processes and synchronization which allow the expression of the models
in a generic manner.

Our approach aims at combining compositional description with automatic model generation.
The formal specification consists in a labelled transition system and
synchronisation networks, in which both events (messages) and
processes (group members) can be parameterized and built from a
graphical language. 
On one hand, having a well-defined  semantics  made  the specification sound; on the other hand, having a framework based on process algebras and bisimulation semantics made possible to benefit from compositionality for specification and verification \cite{handbook:01}.
Parametric synchronisation
vectors also allow us to envision the modelling of dynamic groups with
members joining or leaving the group.

\paragraph*{Related Work}  
Some work has been done to formally verify  properties in group-based
applications. Some of these verifications deal with safety properties,
while others remain limited to a case study. In
\cite{hu99modelchecking} the formal verification of cryptographic
protocols is proposed. It  used  model-checking tool to verify
confidentiality  and  confidentiality properties. 
Model-checking was also used to verify behavioural and dependability
properties \cite{Mas04}. The authors adopted Markov chains to specify
the studied protocols.  
By using a combination of inductive poofs and  probabilistic model
checking \cite{KN02} verified a randomized protocols. 
In the same way, \cite{layouni-correctness03} used a combination PVS
theorem prover and model-checker based on timed-automata for formal
verification of an intrusion-tolerant protocol. \cite{Ban98} presented
a simple deadlock detection mechanism caused by circular synchronous
group remote procedure calls. In contrast with all these, we limit 
ourselves to apply finite model-checking techniques to abstract semantic 
models. Our pNets semantic model is very helpful in this matter, 
providing us with a very expressive and compact formalism, but where 
the usage of parameters is limited in a way that can be easily 
abstracted to finite instances.

Group-based systems as well as parameterized systems are particular infinite
systems in the sense that each of their instances are finite but the
number of  states of the system depends on one or several
parameters. Among these parameters we can distinguish: data structures
or variables (e.g., queues, counters), number of components involved
in the system, ... Automatic
verification of such systems has to face state explosion problem. A
variety of techniques to alleviate state explosion has been
investigated. We can cite: techniques  based on abstraction \cite
{DBLP:journals/entcs/LesensS97, DBLP:conf/vmcai/ClarkeTV06};
techniques  based on finding network invariants \cite {Emerson96, DBLP:journals/fac/SistlaG99,
265960,DBLP:conf/cav/PnueliS00},
which can (possibly-over) approximate the system with an infinite
family of processes.  
Others  \cite{199468, DBLP:conf/icfem/EmersonTW06} based on finding an
appropriate cut-off value of the parameters to bound the system model.
For automatic verification of infinite-state systems
\cite{DBLP:conf/cav/BouajjaniJNT00,DBLP:conf/tacas/AbdullaDHR07}
propose regular model checking. The approach is based on the idea of
giving symbolic representation in term of regular languages. 
Our work tries to take the best of these approaches: whenever possible, we use
property-preserving abstractions to build very small (abstract) data domains 
for the parameters of the basic processes of our systems; but for parameterized
topologies such abstractions are not generally complete, so we have to use
cut-off strategies as in bounded model-checking.
\medskip

In the following of the paper, Section 2 
 overviews  \pa\ communication model and group concepts, and
introduces a running example. Section 3 presents our
theoretical model and its graphical syntax. Section 4 provides a behavioural model for group
communication and synchronisation. 
Section 5 shows our verification methodology, with experimental
results on state-space generation and verification of properties.

\lstset{  language=Java,
         basicstyle=\small,
         identifierstyle=\textit,
         stringstyle=\ttfamily,
}

\section{The \pa~communication model}
\pa~is an LGPL Java library \cite{Pro03} for parallel, distributed,
concurrent applications.  It is based on an active object model, where
active objects communicate by asynchronous method invocation (called
\emph{requests}) with futures: upon a method invocation on an active
object, a request is enqueued at the remote object's side, and a
future is automatically created to represent the result of the request while the
caller continues its execution. Active objects are mono-threaded and
treat the incoming invocations one after the other, returning a value
for the request at the caller as soon as a request is finished. As
remote invocations and future creation are handled transparently, the
programmer can write distributed applications in a much similar manner
to standard sequential ones. In \pa\ there is no shared memory between
active objects to prevent data race-conditions; consequently,
a copy of the request arguments are transmitted to the remote active objects.

\subsection{\pa\ Groups}

In this paper, we focus on the group communication mechanism offered
by \pa~\cite{baduel07asynchronous}. Groups in \pa\ work as follows: a
group of active objects is a set of active objects that behaves as
follows. First,  a method
invocation on the group results in a remote invocation to all the
members of the group in parallel. Second, a list of futures is automatically
created to receive the results returned by the group members. Groups
are typed as usual objects, and thus invocations to a group are made
transparently, as any  object invocation. This way,
specific primitives for groups are only group creation and management,
and thus code modification to handle group communication is minimal.
In \pa, groups are dynamic in the sense that objects can join or leave
the group at runtime.
The main \pa\ primitives for handling groups are the following:
\begin{itemize}
\item
 \lstinline|Group ProActive.newActiveGroup(String Type)|
creates a new group
 of the type ``\texttt{Type}''.
\item \lstinline|void Group.add(Object o)| adds an object to a group.
\item \lstinline|void Group.remove(int index)| Remove the object at the specified index.
\end{itemize}

\subsection{Synchronisation for \pa\ Groups}
\label{SectSynch}
 For classical active objects, synchronisation occurs as follows: a simple
access to the future representing the result of a request  automatically blocks
until the result is computed, and the future is filled.
 For a group invocation, there is one result by group
member,  those results are stored in a group of futures. Synchronizing
on  a group of futures is more complex, here are 3 
synchronization primitives of \pa:
\begin{itemize}
\item \lstinline|void ProActiveGroup.waitAll(Object FutureGroup)| blocks until all the
 futures of the group  return.
\item \lstinline|void ProActiveGroup.waitN(Object FutureGroup, int n)| waits until n
 futures are returned.
\item \lstinline|Object FutureListGroup.waitAndGetTheNth(Object FutureGroup, int n)| waits for the result 
   from the n-th member and returns it.
\end{itemize}

\subsection{Example}
To illustrate group communication, we
consider an application synchronising meetings, it
consists of a master initiator and several  clients that
contain the agendas of the participants. 
The
initiator suggests a date to all participants that reply whether they
are available or not.  For this, we define a class
\lstinline|Participant|:
\begin{lstlisting}
public class Participant {
    Booolean suggestDate(Date d) { ...  }
    Boolean validate() { ...  }
    void cancel()  { ...  }
}
\end{lstlisting}

The following code can be implemented by the initiator to
coordinate the meeting:
\begin{lstlisting}
public static void main(...) {
....
// group creation
  Participant participants=ProActive.newActiveGroup("Participant");
  ...
// we populate the group by adding one or several element 
  participant = ProActive.newActive("Participant",null);
  participants.add(participant);
  ...
  while (true) {
// then we sugggest a date to all members simply by:
  Object answers = participants.suggestDate(date);
  ...
// collateResults gets the result and provides an overall result,
// e.g. returns true if all futures are true
  if (collateResults(answers, ProActiveGroup.size(participants))) {
          Object f=participants.validate();  // validate the meeting
          waitAll(f); // waits until everybody acknowledged validation
          }     
  else
          participants.cancel(); // cancel the meeting
    ... }
}
\end{lstlisting}

\begin{figure}[ht]
\begin{center}
\includegraphics[width=8cm]{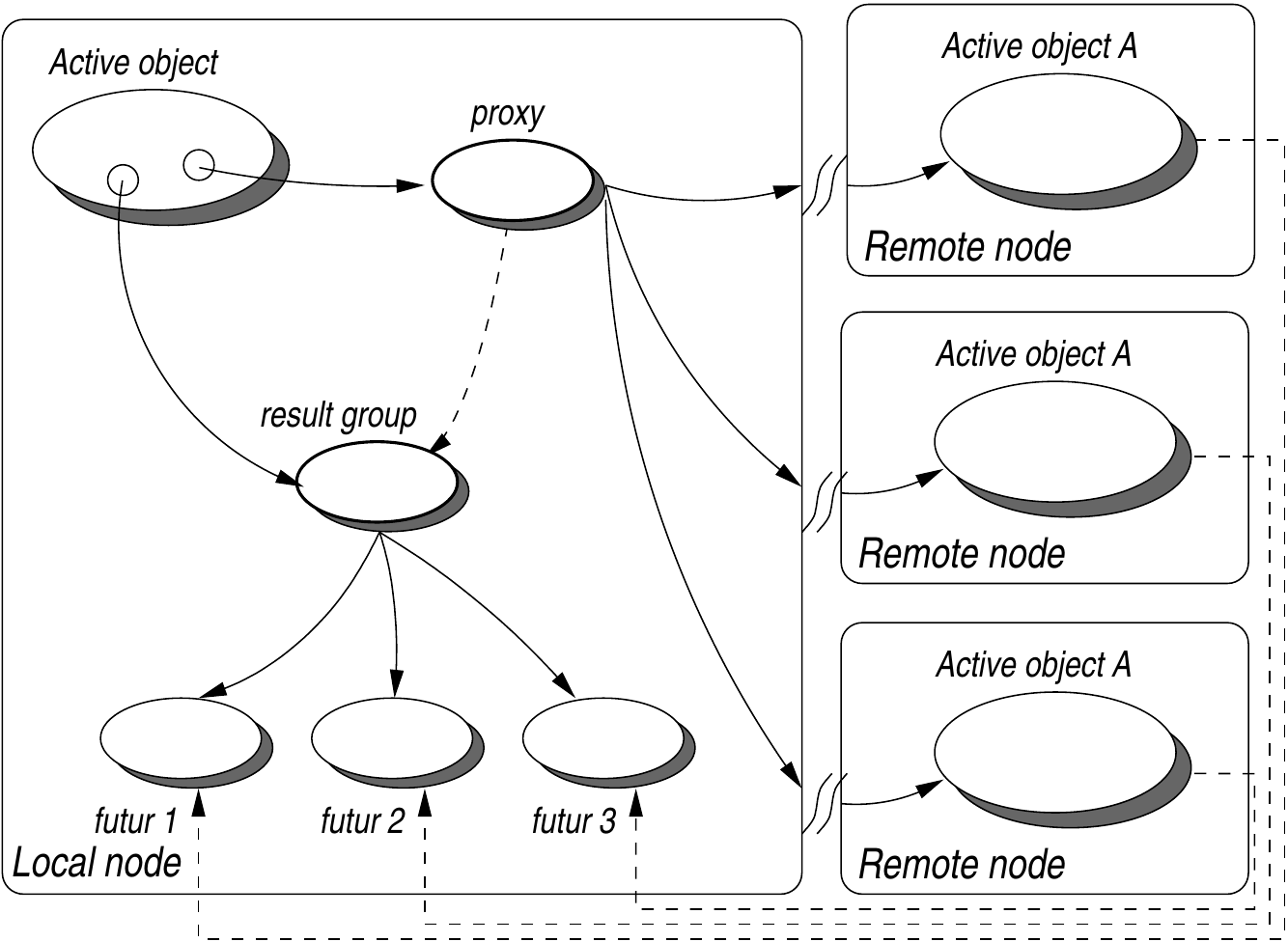}
\end{center}
\caption{\label{Modegroup} Asynchronous and remote method call on group}
\label{fig:inf}
\end{figure}

This example illustrates well the different mechanisms of group
management and communication: first an empty group is created
(\lstinline|newActiveGroup|), then it is populated by several
Participant objects. Thus when the initiator invokes
\lstinline|suggestDate| on the group, this broadcasts a
meeting request to all the members. Then the members reply, which
fills the futures contained in the group of futures
\lstinline|answers|. The local method \lstinline|collateResults| 
synchronises the returns from all these invocations.  Validate or cancel is broadcasted
to all the group members depending on the result of the preceding
step. To illustrate more synchronization mechanisms, the initiator waits until all participants
acknowledge the validation.
A possible implementation of the \lstinline|collateResults| method is the following:
\begin{lstlisting}
boolean collateResults(Object ans, int size) {
  boolean result=true;
  for (int i=0 ; i < size ; i++) {
   if (!ProActiveGroup.waitAndGetTheNth(ans,i)) result = false;
 }
 return result;
}
\end{lstlisting}

Fig. \ref{Modegroup} illustrates the mechanism of group
communication as implemented in \pa. A
method call to a remote activity goes through a proxy, that locally
creates  ``future'' objects, while the request goes to the remote
request queues.

\section{Theoretical Model}

In \cite{BBCHM:article2009}  we have proposed a formalism to represent
the behavior of distributed applications. Behavior of complex systems
can be represented  hierarchically by  composition of classical LTSs
\cite{Milner89}.  Those LTSs are composed using synchronisation
Networks (Net) \cite{Arnold94,Arnold01} so that the synchronisation
product generates  a LTS which can be used at the higher level of hierarchy. Finally the behavior of the system can be expressed by a global LTS. We have also shown that this model can be used as an intermediate format to check  behavioral  properties like temporal ones.

To encode both  families of processes and  data value passing
communication  LTSs and Nets are enriched with parameters~\cite{CansadoM08}. Parameters can be used as communication arguments,
in state definitions, and in synchronisation operators. This enables
compact and generic description of parameterized and dynamic topologies.
In the following we recall  definitions of the {\it parameterized Networks of synchronised automatas (pNets)}  as given in \cite{BBCHM:article2009}. We start by giving the notion of parameterized actions.

\begin{Definition} {\bf Parameterized Actions.} Let $P$ be a set of
 names, $\mathcal{L}_{A,P}$ a term algebra built over $P$,  including
at least a distinguished sort $A$ for actions, and a constant action $\tau$.
We call $v \in P$ a parameter, and $a \in
\mathcal{L}_{A,P}$ a parameterized action, $\mathcal{B}_{A,P}$ is the
set of boolean expressions (guards) over $\mathcal{L}_{A,P}$.
\end{Definition}

\begin{figure*}[!]
\begin{center} 
 \includegraphics[width=14.5cm]{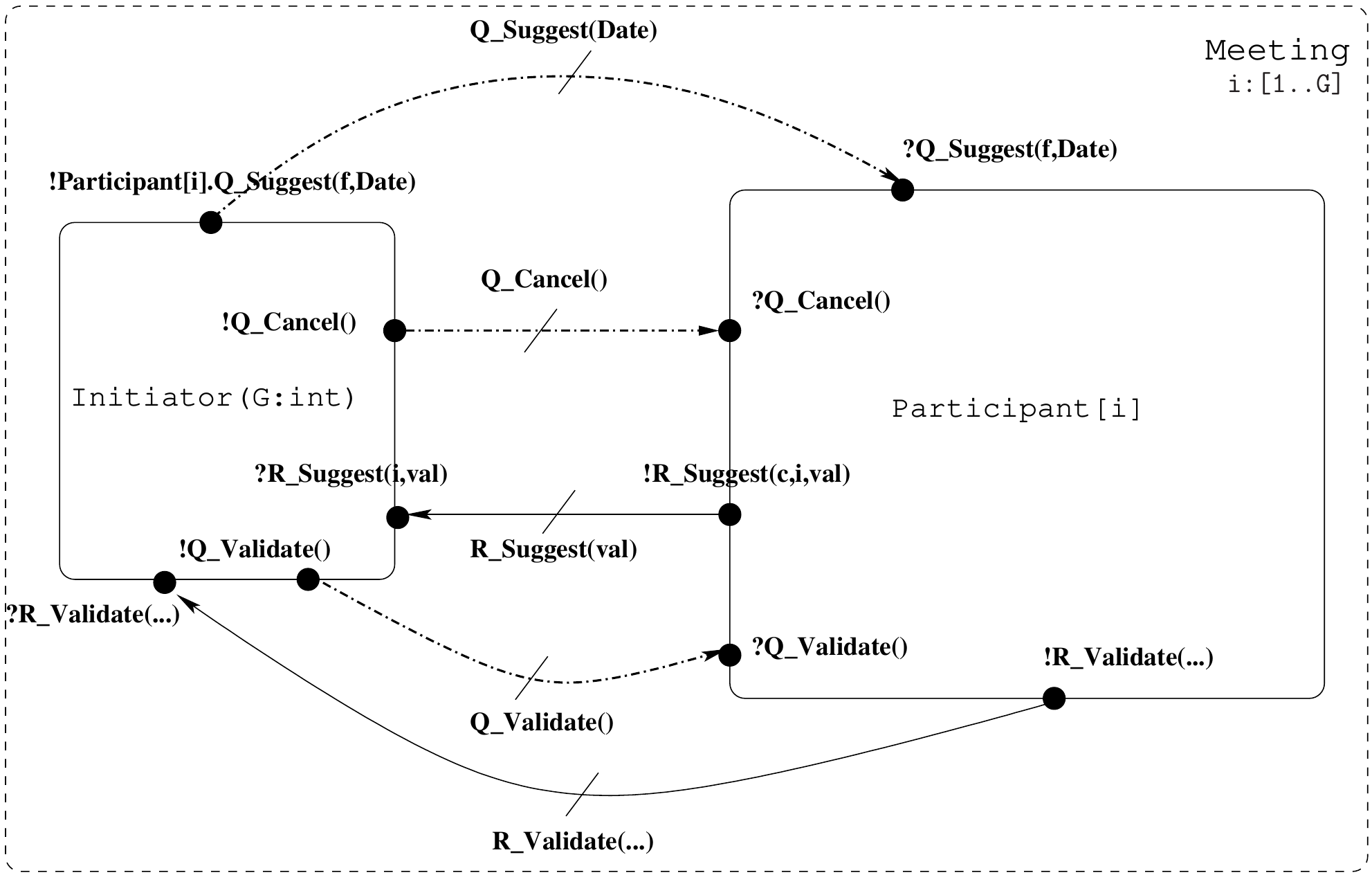}
\caption{Graphical representation of a parameterized network}
\label{figure:MeetNet} 
\end{center}
\end{figure*}

$A$ describes the possible actions representing interactions between
processes. Main actions of our system are illustrated in bold fonts in
Figure~\ref{figure:MeetNet}.  The typical shape of an action is
\textbf{!Participant[i].Q\_Suggest(f,Date)} for a message
\textbf{Q\_Suggest} sent to the member number \textbf{i} of the process
family \textbf{Participant}. \textbf{f} and \textbf{Date} are the
message parameters, here \textbf{f} is the future for the
request, and \textbf{Date} the request parameter. \textbf{!} indicates
an emission, and \textbf{?} a reception. In most cases the
destination of the message can be inferred by the context, and in the
figure by the destination of the arrows, in that case, the actions
look like \textbf{?Q\_Cancel()}.



\begin{Definition}{\bf pLTS}.
\label{pLTS}
A parameterized LTS is a tuple
$\langle P,S,s_0, L, \to\rangle$ where:
\begin{itemize}
\item[$\bullet$]
$P$ is a finite set of parameters,  from which we construct the term
algebra $\mathcal{L}_{A,P}$,

\item[$\bullet$]
$S$ is a set of states; each state $s \in S$ is
associated to a finite indexed set of free variables $\fv(s) =
\tilde{x}_{J_s} \subseteq P$,

\item[$\bullet$]
$s_0 \in S$ is the initial state,

\item[$\bullet$]
$L$ is the set of labels, $\to$ the transition relation $\to \subset S \times L \times S$

\item[$\bullet$]
Labels have the form
$l = \langle \alpha,~e_b,~\tilde{x}_{J_{s'}}\bs:= \tilde{e}_{J_{s'}}\rangle $ such
that if $s \xrightarrow{l} s'$,
then:
\begin{itemize}
\item [-] $\alpha$ is a parameterized action, expressing a combination of
inputs $\symb{iv}(\alpha) \subseteq P$ (defining new variables) and outputs $\symb{oe}(\alpha)$
(using action expressions),
\item [-] $e_b \in \mathcal{B}_{A,P}$ is the \emph{optional} guard,
\item [-] the variables $\tilde{x}_{J_{s'}}$ are assigned during the
transition by the \emph{optional} expressions $\tilde{e}_{J_{s'}}$
\end{itemize}

with the constraints:
$\fv(\symb{oe}(\alpha)) \subseteq \symb{iv}(\alpha) \cup \tilde{x}_{J_s}$ and
$\fv(e_b)\cup \fv(\tilde{e}_{J_{s'}}) \subseteq \symb{iv}(\alpha)\cup\tilde{x}_{J_s} \cup
\tilde{x}_{J_{s'}}$.
\end{itemize}
\end{Definition}

\medskip

We defined  Networks of LTSs called Nets in a form inspired by the {\em synchronisation vectors} of Arnold and Nivat \cite{Arnold94}, that we use to synchronise a (potentially infinite) number of processes. The Nets are extended to  pNets such that  the holes can be indexed by a parameter, to represent (potentially unbounded) families of similar arguments.

\begin{Definition}
\label{pNet}
A {\bf pNet} is a tuple $\langle P,pA_G,J,\tilde{p}_J,\tilde{O}_J,\overrightarrow{V}\rangle $ where:
$P$ is a set of parameters,
$pA_G \subset
\mathcal{L}_{A,P}$ is its set of (parameterized) external actions, $J$ is a finite set
of holes, each hole $j$ being associated with (at most)
a parameter $p_j\in P$ and with a
sort $O_j \subset \mathcal{L}_{A,P}$.
$\overrightarrow{V} = \{\overrightarrow{v}\}$ is a set of
synchronisation vectors of the form:
$\overrightarrow{v}=\langle a_g, \{ \alpha_{t_i}\}_{i \in I, t\in B_i} \rangle $
such that:
$I \subseteq J \land B_i \subseteq \mathcal{D}om(p_i) \land \alpha_{t_i} \in O_i \land  \fv(\alpha_{t_i}) \subseteq P$  
\label{pNet} 
\end{Definition}

Each hole in the pNet has a parameter
$p_j$, expressing that this ``parameterized hole''
corresponds to as many actual processes as necessary in a given
instantiation of its parameter. In other words, the
parameterized holes express {\sl parameterized topologies} of processes
synchronised by a given Net.
Each parameterized synchronisation vector in the pNet expresses
a synchronisation between some instances ($\{t\}_{t\in B_i}$) of some
of the pNet holes ($I \subseteq J$).
The hole parameters being part of the
variables of the action algebra, they can be used in communication and
synchronisation between the processes.

Figure \ref{figure:MeetNet} gives an illustration of a graphical
representation of a parametrized system in our intermediate language.
It shows a meeting system with a single initiator and an arbitrary
number of participants. The parameterized network is represented by a
set of three boxes, {\sc Initiator} and {\sc Participant} boxes inside
{\sc Meeting} box (hierarchy). Each box is surrounded by labelled
ports encoding a particular Sort (sort constraint $pA_G$) of the
corresponding pNet.  The box will be filled with a pLTS or another
pNet (see Fig. \ref{MeetFull}) satisfying the Sort inclusion condition
($L \subseteq pA_G$). The ports are interconnected through edges for
synchronization. 
 Edges are translated to synchronisation vectors.  In previous works we only had single edges with simple arrows
having one source and one destinations, which were translated into
synchronisation vectors of the form
 \texttt{(R\_Validate(),!R\_Validate(),?R\_Validate())} expressing a
 rendez-vous between actions \texttt{!R\_Validate()} and
 \texttt{?R\_Validate()}, visible as a global action
 \texttt{R\_Validate()}. Next section details synchronisation vectors
 for the multiple arrows we use in our example.

\section{Behavioural Model for \pa~Groups}

In \cite{PMDJO:2004} we presented a methodology for generating
behavioural model for \pa\ distributed applications, based on static
analysis of the Java/\pa\ code.  This method is composed of two steps:
first the source code is analysed by classical compilation techniques, with a special attention to tracking references to remote objects in the code, and identifying remote method calls. This analysis produces a graph including the method call graph and some data-flow information. The second step consists in applying a set of structured operational semantics (SOS) rules to the graph, computing the states and transitions of the behavioural model. 

The contribution of this paper is to extend our previous with 
support for group communication and complex synchronizations related to group communication.

The behavioural model is given as a pNets, which we use as an
intermediate language. We express here the semantics of group
communication in this intermediate language and show how behaviour of
application including group communications with various synchronisation
policies can be expressed.

\subsection{Modeling complex synchronisations} 
\label{ModelMult}
\begin{figure*}[!]       
\begin{center}             

 \includegraphics[width=10cm]{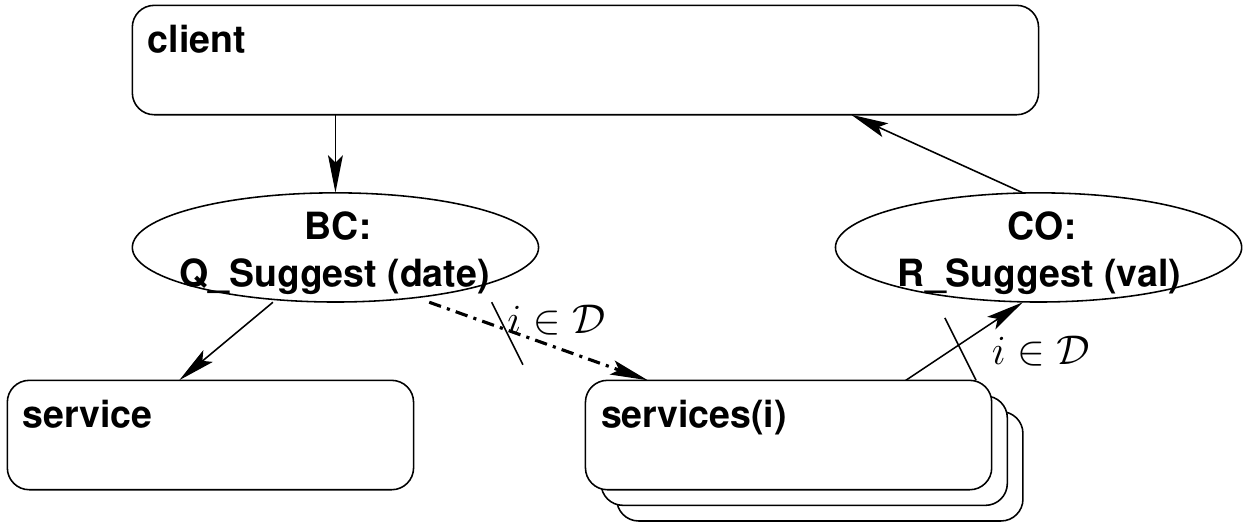} 

\end{center} 
\caption{\label{Synchro} Graphical representation of broadcasting operator}   
\end{figure*}

In order to encode the simultaneity of  several message reception/sending, we use a particular kind of proxy and N-ary synchronisation vectors. 
In Fig. \ref{Synchro} we give a graphical notation for two operators,
the ellipse on the left shows a broadcasting operation, and the one on
the right show a collection operation.

The first operator
that is in charge of  broadcasting  requests to multiple
processes. It is represented by an ellipse with one link
arriving  from a process, and a set of link departing from the
ellipse. The incoming action is triggered as the same time as all the
outgoing ones: in the example the output of the client is triggered at
the same time as the input in the service on the left, and the input
in all the services on the right (the dotted arrow
denotes a multiple link).
 We extend parameterized vectors to support the multicasting communication.

For broadcasting, we introduce the \texttt{BC} operator to encode a
family of synchronized processes. The vector \(< Q\_suggest,
!Q\_suggest(date), BC ~ i\in\mathcal{D}.\: ?services[i].Q\_suggest(date),
?service.Q\_suggest(date)>\) indicates the synchronisation between one
instance of the network 1 (client), a given number of network 2 (services),
and another service process. The synchronisation is an observable
action labeled $Q\_suggest$.
The parameter $i$  ranges in the domain $\mathcal{D}$. For instance, if $\mathcal{D}=[0..1]$, then the vector is expanded to: \\
{\footnotesize \(< Q\_suggest, !Q\_suggest(date), ?services[0].Q\_suggest(date),?services[1].Q\_suggest(date), ?service.Q\_suggest(date)>\)}.

The operator on the right side collects communications: it
synchronizes \emph{one} of its input with its single output.  For
encoding such a synchronisation, we introduce the \texttt{CO} operator
to encode a set of synchronisation vectors. The vector \(<
R\_suggest(val),?R\_suggest(i,val), CO i\in\mathcal{D}.\: !services[i].R\_suggest(val)
>\)
indicates  the synchronisation between a \texttt{R\_suggest} action in the
network 1 (client) and an output of one of the network 2 (services). For instance, with $\mathcal{D}=[0..1]$, this vector is expanded to several vectors:\\
\(< R\_suggest(val), ?R\_suggest(0,val), !services[0].R\_suggest(val), *>\)\\
\(< R\_suggest(val), ?R\_suggest(1,val), *, !services[1].R\_suggest(val)>\)

Those two synchronisation mechanisms will be further illustrated in
the encoding of the example.

\subsection{Modeling the Example} 

\begin{figure*}[!]        
\begin{center}             
 \includegraphics[width=19cm,angle=90]{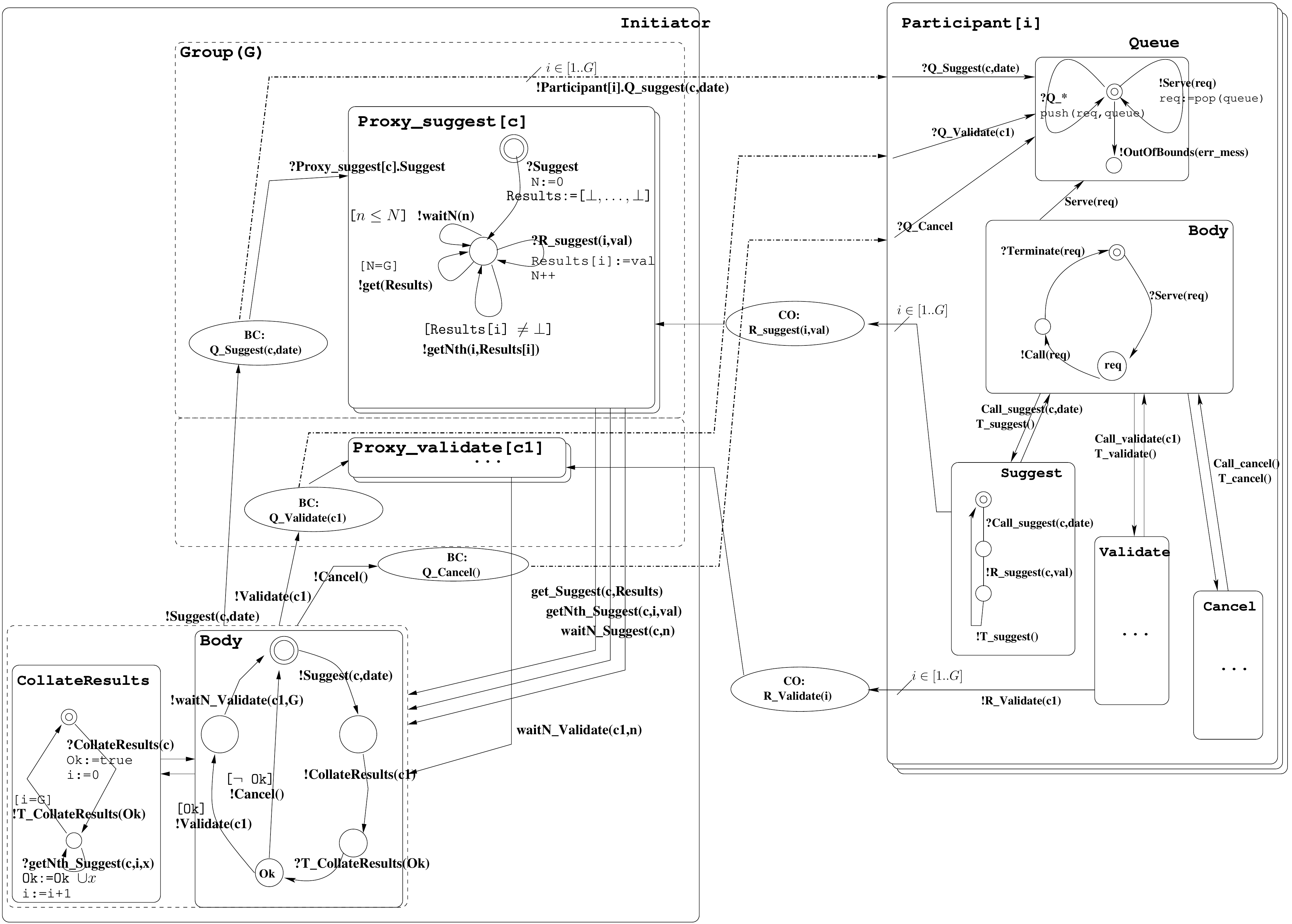}
\end{center} 
\caption{\label{MeetFull} Model of a communication by broadcasting}   
\end{figure*}

We describe now the behavioural model for our example
application, especially focusing on the modeling of group proxies, and
the  communications involving groups. The full model for our example is shown in
Fig. \ref{MeetFull}. The model is split into two parts interconnected
by parameterized synchronization vectors.
\begin{itemize}
\item \emph{The initiator} encodes a client side behaviour. The
  Initiator contains a body encoding an abstraction of the functional
  code, and the group proxies. For each remote method call in the
  Initiator code there is a parameterized group proxy, representing an
  unbounded number of future proxy instances. The body repeatedly
  suggest a date and either cancel or validate depending on the
  answers.

\item \emph{The participants} encodes the server side behaviour.
  They are modelled by an indexed family of processes, each representing
  the behaviour of one element of the group, with its request queue,
  its body serving requests one after the other in a FIFO order, and
  the code of its local methods.
\end{itemize}

A {\bf Proxy} pNet (box) is created for each remote method invocation.
The Proxy is
indexed by the program point ($c$) where the method is called.
 The
{\bf Proxy} pNet models the creating and the management of the group
of futures: Once the group of future is created, futures can be
received one after the other, and each already received future can be
accessed. It is also possible to wait until $N$ answers are
received.

For each remote method call of the Initiator, a broadcast node,
synchronizes the sending of the method call by the initiator body, the
initialisation of the corresponding future, and the reception of the
request message in the queues of each of the participants in the group.

Concerning the user code, the {\bf Body} boxes in Fig. \ref{MeetFull} represent the behaviour of
the main method of each active object, again on the form of a
pLTS. The code for each method (e.g. \textbf{Validate}) is also expressed by a
pLTS, and triggered when serving the corresponding request, or by
direct invocation like \textbf{collateResult}.
Each of them is either obtained by source code analysis, or provided by
the user.

As it is the only object to act as a server, the participant has a
\textbf{Queue} box. The corresponding pLTS  encodes a FIFO queue of
request that is accessed by the participant's body, and filled when
the initiator sends a request. The queue can be given a maximum length
and raise an error if it is overflowed.


\subsection{Variations on group synchronisations}
\label{Var-Syn}

\pa~provides various primitives (see Section \ref{SectSynch}) allowing
the programmer to control explicitly the synchronization of
asynchronous methods calls by waiting the incoming replies. The
network {\bf Proxy\_suggest} in Fig. \ref{MeetFull} specifies three
kinds of these primitives: {\it waitAll}, {\it waitN} and {\it
  waitAndGetTheNth}. Those three primitives show the different
synchronisations that our group proxies can express: counting the
number of returned objects, or returning a specific result.
They are encoded very naturally using a table of received results, and
the number \texttt{N} of results already returned. Those information
are updated when receiving messages from a   \emph{collection} (\texttt{CO}) of different results as
explained in Section~\ref{ModelMult}.
Additionally to those primitives, one could also use a
\textit{waitOne} primitive waiting for one result, no matter of which
it is; this primitive could be encoded with a little more effort by
our proxy, but we do not present it because it is not used in our
example and we believe it is less crucial than the others.
\textit{waitOne} is useful in the case several workers perform
the same task, and only one result is necessary. 



\section{Verification and Results}

\begin{figure*}[ht]         
\begin{center} 
\fbox{\begin{minipage}{.99\textwidth}
\noindent\small
Abstract data domains:  
 $Group$ index: G $\in$ [0..2],
$Q\_Suggest$ argument: $data \in \{D1,D2\}$,
$Q\_Suggest$ result: $bool$ 
\smallskip

Observed sorts: \\
Initiator sort: $\{Q\_Suggest(data), Q\_Validate(), Q\_Cancel(),
R\_Suggest(index,bool), R\_Validate(index), $ \\
\ \hspace{3cm} $T\_CollateResults(bool)\}$ \\
Participant sort: $\{Q\_Suggest(data), Q\_Validate(), Q\_Cancel(),
R\_Suggest(bool), R\_Validate(), Error()\}$ \\
ParticipantGroup sort: $\{Q\_Suggest(data), Q\_Validate(), Q\_Cancel(),
R\_Suggest(index,bool), R\_Validate(index), $ \\
$Error()\}$ \\
System sort: $\{Q\_Suggest(data), Q\_Validate(), Q\_Cancel(),
R\_Suggest(index,bool), Error(),  $ \\
$T\_CollateResults(bool)\}$ \\

\begin{tabular}{|p{4.5cm}|r|r|r|r|r|r|}
\hline
\multirow{2}{*}{\textbf{Subsystem}}&\multicolumn{2}{c|}{\textbf{brute force}}&\multicolumn{2}{c|}{\textbf{minimized}}&\textbf{gen. + min.}\\

              &\textbf{nb states}&\textbf{nb transitions}&\textbf{nb states}&\textbf{nb transitions}&(seconds)\\ \hline
Single Participant & 1 801 & 5 338 & 90 & 376 & 8.2 \\
\hline

Initiator & 3 163 & 152 081 & 54 & 1 489 & 11.3\\ \hline

Full system: & & & & & \\
with 3 participants, queue[1] & 85 213 & 839 188 & 178 & 489 & 17.9 \\
with 3 participants, queue[2] & 170 349 & 1 646 368 & 458 & 1 284 & 406.0 \\ \hline

\end{tabular}

\begin{tabular}{|p{4.5cm}|l|r|c|c|}
\hline
With Distributed generation&{\textbf{generation}}&{\textbf{Total Time}}&{\textbf{States/Transitions}}&{\textbf{States/Trans}}\\ 
&{\textbf{algorithm}}&&&{\textbf{(minimized)}}\\ 
\hline

Full system with 3 participants & brute force & 6'45'' &
170 349 / 1 646 368 & 458 / 1 284 \\
 (8x4 cores)& tauconfluence &   30' & 5591 / 14 236 & 458 / 1 284\\ \hline

Group of 2 participants   & brute force & 11'32"  & 13 327 161 / 48 569 764  & 4 811 / 24 588\\
(15x8 cores)  & tauconfluence & 1150'55'' & 392 961 / 1 354 948  & 4 811 / 24 588 \\ 
\hline

Group of 3 participants  & tauconfluence & - & {\sl Out of memory} & -\\ 
 (15x8 cores)&  &  & {\sl estimate $\geq 10^{11}$ states}  & \\\hline

\end{tabular}

\end{minipage}
}
\end{center}
\caption{\label{GeneratedSizes} Size of the generated state spaces for
different sub-systems of our example}   
\end{figure*} 

In principle, the steps for designing and validating a distributed application with our approach are:
\begin{enumerate}
\item Specify the structure and the behaviour of the application, in terms of active objects (or components). 
We provide editors for distributed components in the Vercors
platform; specific component interfaces exist for group communication.
Alternatively one could imagine tools for static analysis of Java/ProActive code, that would provide a similar abstraction of the system.
\item Generate a pNet model, following the approach in the previous section. We plan to have tools automatizing this step in a near 
future, integrated in the Vercors platform.
\item Write user requirements, in the form of logical formulas in some temporal logic dialect (most action-based logics will be suitable).
\item Use a model-checker to check the validity of theses formulas on the generated model. Currently only finite-state model-checkers
are capable to analyse our models. This means that the parameterized pNets have to be instantiated first to a finite system, and that 
the formulas have to be instantiated accordingly.
\end{enumerate}

The reader acquainted with model-checkers will have guessed that such models are severely exposed to state explosion. It is very 
important here to observe two facts: 
First we only work with an abstraction of the system. We use finite abstractions of data-values
in the description of data domains, and we only expose (and observe) the events that are useful for the properties.
Secondly, we make use as much as possible of the congruence properties
of our semantic model: we build the state-space in a hierarchical
manner, often minimizing partial models using branching bisimulation before
building their products. But this strategy has limits, and sometimes
it is better to build the state-space of a subsystem under the
constraints of its environment, avoiding unnecessary complexity; this
is illustrated in our case-study by the ``Participant group'' that has
by itself a very high state complexity, of which only a small part is
used by the ``Initiator'' client.

In Figure \ref{GeneratedSizes} we give figures obtained on our example. The systems in the first 4 lines of the table
have been computed on a Fedora 10 box, with 2 dual-core Intel
processors at 2.40 GHz, with a total of 3.8 Gbytes of RAM. The source
specification was written in the intermediate format Fiacre \cite{BERTHOMIEU:2003, BERTHOMIEU:2008},
and the state space generated using CADP version 2008-h. 
The systems in the last part of the table have been computed on a
cluster with 15 nodes, each having 8 cores and 32 Gbytes of RAM. We
have been using the Distributor tool of CADP for distributed
state-space generation, with or without on-the-fly reduction by
tauconfluence \cite{GaravelSerwe2006}; the distributed state space has to be merged
into a single state space before minimization and model-checking.
The execution times in this part include the deployment of the application,
the distributed generation, the merging and the minimization of the resulting state-space.
A cell with a ``-'' means that the computation did not terminate.

The main lesson from this experiment is that intermediate systems will often cause the main 
bottlenecks in the system construction. Here, an unconstrained model for a group of 3 participants 
is already too big to be computed on a single desktop machine. By contrast, computing the behavior 
of such a group in the context of a specific client is feasible (here
the model of the full system with 3 participants remains reasonably small).
Generating the state-space in a distributed fashion gives us the
capability of handling significantly larger models. On-the-fly
reduction strategies are useful too, but to a certain point only,
because it may involve local computations that require large local
memory space themselves. 
In our tests the generation of the model of a
group with 3 participants failed: we estimated that the brute force model
has approximately 125 billiards of states (this would require some 12
Terabytes of distributed RAM, 25 times more than our full
cluster). But even using on-the-fly reduction by tauconflence, local
computations caused an out-of-memory failure.


\paragraph{Proving properties} We give here examples of functional
behavioural properties that we checked on various scenarios.
For this, we have built the global synchronisation product of the system, with 3 Participants in the group (the number of participants does not change the results), and with the size of requests queues instantiated to 1 or 2 depending of the cases.

For expressing the properties, we could  use any of the logical
languages provided within the CADP tool suite, including LTL, CTL, or
specification patterns \cite{Dwyer98propertyspecification}. In
general, we use the
regular alternative-free $\mu$-calculus formalism, which is a powerful
modal logic, nicely expressing action sequences as regular
expressions; it is the native logics of the model-checker.
We have checked the following formulas:
\begin{enumerate} 
\item \(<True*.Error> True \) : in the system with queue of length 1, the queues can signal an Error.
\item \([True*.Error] False \) : in the system with queue of length 2, the queues never signal an Error.
\item \(< True*.R\_suggest(i,b) > True  \) : some paths lead to
  a response to the \texttt{suggest} request.
\item \( <True*.T\_CollateResult(false)> True \) :  the collection of results by the Initiator
  can return \texttt{false}.
\item \(After\: !Q\_Suggest(id)\: Eventually\: !Q\_Cancel() \lor
  !Q\_validate() \) : inevitable reachability of either a
  validation or a cancellation after a date has been suggested. 
This formula is written in the specification patterns formalism, and expresses correct progress of the system. 
\end{enumerate}
Properties 1. and 2. are checked on two different models, with
different size of the queue. They prove that a bounded queue of length
2 is required and sufficient to ensure the correct operation of the
system. The Error action in the queue of a participant signals that a
request is received in a state where the Queue is already full.

Properties 3. and 4. check the reachability of some possible events; technically, property 3 has to be checked for each possible values of parameters i and b, because the $\mu$-calculus logic is not parameterized.

Property 5. expresses  the correction of the (first iteration of
the) behaviour of the system: in response to a suggest request, we
guarantee that the initiator sends  either a validation or a cancellation message.

It is interesting to discuss the tools available for exploring and debugging the generated systems. In addition to the model-checking and minimization engines, we have used tools for:
\begin{itemize}
\item exploring interactively the generated behaviour at the level of its Lotos representation (OCIS)
\item displaying graphically the generated LTS (BCG\_EDIT) 
\end{itemize}

\begin{figure*}[h]        
\begin{center}   
 \includegraphics[width=15.5cm]{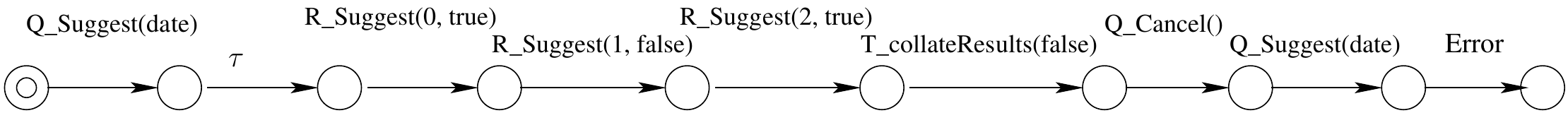}
\end{center} 
\caption{\label{Path} Path containing the Error action}   
\end{figure*}

Consider formula 1 that checks reachability of action Error. In
addition to a ``True'' result, the
model-checker produces a trace
illustrating the reachability from the initial state, as shown in
Figure \ref{Path}.  The trace consists in a full cycle through the
system behaviour, from the initial state to state 6 and action
``Q\_cancel()''. Then, because we do not wait for the  return of the
Cancel requests,  one of the Participants can still have a Cancel
request pending in its queue when the Initiator sends the next
Suggest request, which leads to an Error.  The BCG\_EDIT tool can
display the sequence of Figure~\ref{Path}. A finer trace
showing  internal interactions and allowing  user-driven
guidance of the system can be obtained with the OCIS tool.

\section{Conclusion} 
In this  paper we have sketched models for specifying and verifying
the correct behaviour of group-based applications.  Our parameterized
models enable the finite representation 
of groups of arbitrary size, and express the communication with such
groups, together with the associated synchronizations. For our
modelling, we focused on
 the \pa\ library; nevertheless these models can be applied to other middlewares involving collective communications. 
Our parameterized  models are supported by  model checking tool. 
Besides they  are hierarchical labelled transition systems,  
therefore suitable for analysis with verification tools based on bisimulation semantics. 

Our main contribution  is to provide a
behavioural semantic model  for group communication  applications. It
allows the application programmer to prove the correctness of his/her behavioral
properties, and for instance  detect deadlocks \cite{Ban98}. 
We have illustrated our approach on an example application,  generated
the corresponding model, and proved several 
properties ensuring the correct behaviour of the example.
The size of the generated system and the proven properties show that,
if the system is entirely known at instantiation time, we are able to
prove non-trivial properties on examples of a reasonable size.

\paragraph{Towards dynamic groups}
A nice perspective of this work is the verification of groups with
dynamic membership. The \pa\ middleware allow active objects to join and
leave a group during execution. This way the application can adapt
dynamically in the case new group members are necessary to perform a
complex computation, or systematically when new machines join the network. 
The use of pNets will facilitate the specification of dynamic groups
thanks to the support for parameterized processes and synchronisation
vectors. 

\bibliographystyle{eptcs}

\bibliography{Oasis}

\end{document}